\documentclass[pra,showpacs,12pt]{revtex4-1}
\usepackage{graphicx}
\usepackage{float}

\begin{document}

\title[]{Quantum correlations and entanglement in a model comprised of a short chain of nonlinear oscillators}

\author{J. K. Kalaga}
 \altaffiliation[]{Quantum Optics and Engineering Division, Faculty of Physics and Astronomy, University of Zielona G\'ora, Prof.~Z.~Szafrana 4a, 65-516 Zielona G\'ora, Poland}
\author{A. Kowalewska-Kud{\l}aszyk}%
 \altaffiliation[]{Nonlinear Optics Division, Faculty of Physics, Adam Mickiewicz University, Umultowska 81, 61-614 Pozna\'n, Poland}
\author{W. Leo\'{n}ski}%
 \altaffiliation[]{Quantum Optics and Engineering Division, Faculty of Physics and Astronomy, University of Zielona G\'ora, Prof.~Z.~Szafrana 4a, 65-516 Zielona G\'ora, Poland}
\author{A. Barasi\'{n}ski}%
 \altaffiliation[]{Quantum Optics and Engineering Division, Faculty of Physics and Astronomy, University of Zielona G\'ora, Prof.~Z.~Szafrana 4a, 65-516 Zielona G\'ora, Poland}

\date{\today}
\begin{abstract}
We discuss a model comprising a chain of three Kerr-like nonlinear oscillators pumped by two modes of external coherent field. We show that the system can be treated as nonlinear quantum scissors and behave as a three-qubit model. For such situation different types of tripartite entangled states can be generated, even when damping effects are present in the system. Some amount of such entanglement can survive even in a long-time limit. The flow of bipartite entanglement between subsystems of the model and relations among first-, second-order correlations and the entanglement are discussed.
\end{abstract}


\maketitle
\section{Introduction}

Quantum correlations, including quantum entanglement, seems to be one of the most intriguing problems of contemporary physics.  Such correlations play a crucial role not only in searching for the answers on the most fundamental questions concerning laws ruling quantum reality but also in more practical applications as those related to the quantum information theory. As the bipartite entanglement and other forms of quantum correlations seem to be well understood, such correlations in systems containing three and more subsystems still need thorough investigation. Therefore, finding physical models which would be helpful in the research in this field, and from other side, to be general enough to be applied in various physical realizations, became very important problem.

Good candidates for such models are those involving quantum Kerr-like nonlinear oscillators. The effective Hamiltonians describing such systems involve terms with third  order susceptibilities (Kerr-like nonlinearities). Quantum Kerr-like models are widely discussed in numerous applications. For instance, they were considered as a source of non-Gaussian motional states of trapped ions \cite{SVL11}, source of the superposition of coherent states \cite{MTK90,PT92}, and were discussed in a context of the Bell's inequality violations \cite{SJH07}. Moreover, Kerr-like oscillatory models were subject of numerous papers related to the quantum chaos problems (for instance see \cite{L96,*SGM06,*KKL08,*KKL09,*SG12,*KKL12,*SCK13,*GSCK13}).

What is important, models described by the Hamiltonians involving Kerr-type nonlinearities can be found in various, not necessairy optical,  physical systems. Such nonlinearities are applied in description of nanomechanical resonators and various optomechanical systems \cite{BJK97,J07,SMW08,LMG10,R11,WGL15}, boson trapped lattices \cite{WMW09,BLS14,IMP15}, Bose-Einstein condensates \cite{PLK13}, Bose-Hubbard chains \cite{O15a,*O15b,IMP15,PMT15} or in circuit QED models \cite{BGB09,RTM09} (also involving superconducting systems \cite{HSS11,DPB11}).

One of the key problems related to the quantum information theory is quantum state engineering allowing for the finite-dimensional states generation. Such states can also exhibit  such interesting properties as their ability to produce various kind of quantum correlations, including quantum entanglement. Physical systems  involving at least two separate components characterized by the third  order susceptibilities (Kerr-like nonlinearities) are  those  which  allow for  creation of  such quantum states. Although such multimode systems can be found in various physical situations, they are referred to as  Kerr-like couplers -- their evolution is governed by the same effective Hamiltonians as usual optical Kerr couplers (discussion of such optical systems can be found for instance in \cite{KP97,*KP97a,*AP00}, including review paper \cite{PP00}).

 As it was shown in \cite{LM04,*ML06,*KL06,*KLP11}, Kerr-like couplers can be treated as \textit{nonlinear quantum scissors} (NQS) \cite{LK11}. NQS systems exhibit such evolution for which only some limited number of n-photon states is involved.  After such truncation of the wave-function, a coupler playing NQS role, can be treated as 2-qubit \cite{LM04,ML06}, 2-qutrit \cite{KLP11} or qutrit-qubit \cite{KL06} model. It was also shown there that various maximally entangled states can be generated by such systems, including not only usual Bell states but also generalized ones \cite{KLP12}. 

In fact, in NQS models we observe so called photon (phonon) blockade effects. At this point one should mention that such effects were also widely discussed. For, instance they have been observed in systems involving  optical cavity with trapped atom (Caltech experiment \cite{BBM05}),  quantum dot coupled to photonic crystal resonator (Stanford experiment \cite{FFE08}), superconducting qubit coupled to transmission line resonator (ETH-Zurich experiment \cite{LBE11}) or in superconducting circuit (Princeton-NIST experiment\cite{HSS11}). Moreover, quite recently the experiment with microwave cavity coupled to the superconducting qubit  was performed \cite{BCF15}.

Kerr-like oscillators were also discussed in a context of various  special quantum states appearing in the  tripartite systems. For instance, three nonlinear oscillators mutually coupled to each other were used as a source of entangled W states  \cite{SWU06}. It is particularly important because recently, tri and multipartite entanglement becomes one of the most intriguing features discussed in the literature (for instance see \cite{XFS12,*PA13,*WCC15,*O15a,*O15b,*AWT15}). 

The main aim of the present paper is to show how various types of quantum correlations can appear during the evolution of a chain of Kerr-like oscillators externally pumped by  coherent fields. In particular we are interested in generation of tripartite entangled states belonging to various classes. It appears that for such  system, during its evolution, depending on the strengths of the internal and external coupling strengths, GHZ, W and other types of entangled tripartite states can be generated. Moreover, we show that bipartite entangled states are generated not only for the subsystem involving two directly interacting oscillators but also for oscillators interacting  only via the third one. The degree of this entanglement is not fragile to the internal interactions strength. To check how  quantum correlations discussed are fragile for dissipation effects,
the dynamics of a Kerr-type chain is  studied also for the two models of external reservoirs: amplitude and phase damping ones. It appears that in the amplitude damping reservoir in steady state limit, even for relatively large damping constants, some amount of tripartite entanglement can be left in the system.
 
In the literature we can find the discussion concerning problems of relations among the correlation functions (of 1-st and 2-nd order) and entanglement obtained in various quantum systems, such as atomic ensembles in high-Q cavities \cite{SLG11} or some optomechanical systems \cite{SLF12}. It was shown there, that the entanglement is sensitive to the first order coherence between the studied modes and that they are not simultaneously present in the system. The problem of such relation for the tripartite system  will also be discussed in the present paper. We shall also focus on the second order correlation function with reference to the both: two- and tripartite entanglement.

\section{The model and solutions}
\begin{figure}[h]
\begin{center}
\includegraphics[width=0.6\textwidth]{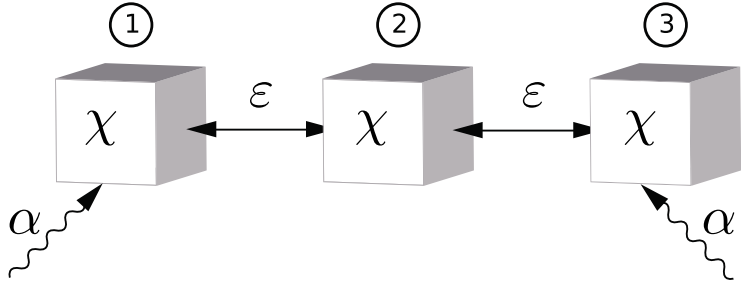} 
\end{center}
\caption{ Scheme of the model. A chain of nonlinear oscillators which are coupled mutually by linear interaction of the strength $\varepsilon$. Two (boundary) oscillators are excited by external coherent field (coupling strength is labelled by $\alpha$).}
\label{fig1}
\end{figure}

We discuss a model of  three identical nonlinear Kerr-like quantum oscillators. They are coupled each other by linear interaction,  and form a chain of oscillators (see Fig.\ref{fig1}) -- one central oscillator and two boundary ones. Moreover, two of them (boundary oscillators) are excited by external coherent field. The system is governed by the following effective Hamiltonian:
\begin{equation}
\hat{H}=\hat{H}_{nl}+\hat{H}_{i}+\hat{H}_{e},
\label{h}
\end{equation}
where
\begin{equation}
\hat{H}_{nl}=\frac{\chi}{2}\left(\hat{a}_{1}^{\dagger}\right)^2\hat{a}_{1}^2+\frac{\chi}{2}\left(\hat{a}_{2}^{\dagger}\right)^2\hat{a}_{2}^2+\frac{\chi}{2}\left(\hat{a}_{3}^{\dagger}\right)^2\hat{a}_{3}^2
\label{H_nl}
\end{equation}
describes free evolution of the oscillators,
\begin{equation}
\hat{H}_{i}=\left( \varepsilon\hat{a}_{1}^{\dagger}\hat{a}_{2}+\varepsilon^{*}\hat{a}_{2}^{\dagger}\hat{a}_{1}\right)+
\left( \varepsilon\hat{a}_{2}^{\dagger}\hat{a}_{3}+\varepsilon^{*}\hat{a}_{3}^{\dagger}\hat{a}_{2}\right)
\end{equation}
represents internal interaction within a pairs of two subsequent oscillators ($1-2$ and $2-3$), whereas
\begin{equation}
\hat{H}_{e}=\alpha\hat{a}_{1}^{\dagger}+\alpha^{*}\hat{a}_{1}+\alpha\hat{a}_{3}^{\dagger}+\alpha^{*}\hat{a}_{3}
\end{equation}
describes interaction with external field. The~operators $\hat{a}_{j}^{\dagger}$ and $\hat{a}_{j}$ $(j=1, 2, 3)$ are bosonic creation and annihilation operators corresponding to the three modes 1, 2, 3, respectively. The~parameter $\varepsilon$ describes the strengths of linear couplings between modes $1-2$ and $2-3$, where we assume that the~both couplings ($1-2$ and $2-3$) are equal each other, and the parameter $\varepsilon$ is assumed to be real \textit{i.e.} $\left( \varepsilon = \varepsilon^{*}\right) $. The strengths of excitations of the oscillators $1$ and $2$ by external field are identical, and are described by  $\alpha$, where $ \alpha=\alpha^{*}$. We deal here with bosonic systems commonly applied in quantum optics. Nevertheless, it is worth noting that non-trivial coupling between boson's degree of freedom, can be also important in the condensed matter physics, where the superconducting state is often observed (for instance see \cite{SD14,*DS14,*SDS14,*DSS14}).

To describe the system's evolution for the case when damping effects are neglected, we shall apply standard Schr\"odinger equation solution method. Therefore, we define the following  three-mode wave-function defined in  $n$-photon Fock states basis:
\begin{equation}
	\label{eq:wave_fun}
\vert\psi(t)\rangle = \sum_{i,j,k = 0}^{\infty} C_{ijk}(t)\vert i \rangle_1 \otimes \vert j \rangle_2 \otimes \vert k \rangle_3\,\equiv\, \sum_{i,j,k = 0}^{\infty} C_{ijk}(t)\vert ijk \rangle,
\end{equation}
where $C_{ijk}$ are complex probability amplitudes corresponding to the states $|ijk\rangle$.

If we assume that external excitation and coupling between oscillators are weak if compare them to the nonlinearity parameters $(\alpha, \varepsilon \ll \chi)$, the system's evolution remains closed within a finite set of $n$-photon states. For such situation the system behaves as \textit{nonlinear quantum scissors} (NQS) ( for instance, see \cite{LT94}) and the review paper \cite{LK11}) and its wave-function is defined in finite-dimensional Hilbert space \cite{MLI01,*LM01}. What is interesting, NQS effect is in fact equivalent to the \textit{photon/phonon blockade} \cite{ISW97,*GWG98,BBM05,LMG10,*DPB11,*MPL13,*MBP14}. 
For the situation discussed here we also assume that system's evolution starts from the vacuum state $|000\rangle$ and hence, it is limited to only eight states $\vert ijk \rangle$ $i, j, k \in \{0, 1\}$. 
In consequence, the truncated wave function can be expressed in following form:
\begin{eqnarray}
	\label{eq:cut_wave_fun}
\vert\psi(t)\rangle_{cut}& =& \; C_{000}(t)\vert 000 \rangle + C_{001}(t)\vert 001 \rangle 
 +C_{010}(t)\vert 010 \rangle + C_{011}(t)\vert 011 \rangle \nonumber \\
 & +& C_{100}(t)\vert 100 \rangle + C_{101}(t)\vert 101 
   + C_{110}(t)\vert 110 \rangle + C_{111}(t)\vert 111 \rangle .
\end{eqnarray} 
Such cutting of  Hilbert space, and hence the wave-function, is related to the fact that the Hamiltonian (\ref{H_nl}) generates unevenly spaced energy levels, where the vacuum and one-photon states are degenerate with their energies equal to zero. For the situation considered here those states are resonantly coupled each other by weak, zero-frequency couplings, and are dominant in the system's evolution. To describe the contribution of two- (and more) photon states we need higher-order perturbative solutions (see \cite{LDT97}).
Then, applying standard procedure  we find equations of motion for our system's dynamics.
As $\vert\psi(t=0)\rangle =\vert 0  0 0\rangle$, these equations lead to the following formulas determining time-evolution of the probability amplitudes:
\begin{widetext}
\begin{eqnarray}
C_{000}(t)&=&\frac{1}{8\alpha^4+2\varepsilon^4}\left(4\alpha^4-2\alpha^2\varepsilon^2+2\varepsilon^4+\alpha^2\left(A_{1}\cos(\omega_{1}t)+A_{2}\cos(\omega_{2}t)\right)\right)\, ,\nonumber \\ 
C_{001}(t)&=&C_{100}(t)=\frac{-i \alpha}{2}\left(\frac{\sin(\omega_{1}t)}{\omega_{1}}+\frac{\sin(\omega_{2}t)}{\omega_{2}}\right)\, ,\nonumber \\ 
C_{010}(t)&=&\frac{\alpha}{8\alpha^4+2\varepsilon^4} \left(-2\varepsilon^3+A_{3}\cos(\omega_{1}t)+A_{4}\cos(\omega_{2}t)\right)\, ,\nonumber \\ 
C_{011}(t)&=&C_{110}(t)=\frac{-i \alpha}{2}\left(\frac{\sin(\omega_{1}t)}{\omega_{1}}-\frac{\sin(\omega_{2}t)}{\omega_{2}}\right)\, ,\\
C_{101}(t)&=&\frac{\alpha}{8\alpha^4+2\varepsilon^4} \left(-4\alpha^3+2\alpha\varepsilon^2+A_{3}\cos(\omega_{1}t)+A_{5}\cos(\omega_{2}t)\right)\, ,\nonumber \\
C_{111}(t)&=&\frac{\alpha^2}{8\alpha^4+2\varepsilon^4}\left(4\alpha\varepsilon+A_{1}\cos(\omega_{1}t)-A_{2}\cos(\omega_{2}t)\right),\nonumber
\end{eqnarray}
\end{widetext}
where two frequencies $\omega_{1}$ and $\omega_{2}$ are defined as:
\begin{eqnarray}
\omega_{1}&=&\sqrt{4\alpha^2+4\alpha\varepsilon+2\varepsilon^2} ,\nonumber \\
\omega_{2}&=&\sqrt{4\alpha^2-4\alpha\varepsilon+2\varepsilon^2} ,
\label{eq9}
\end{eqnarray}
and we have defined the following parameters:
\begin{eqnarray}
A_{1}&=&2\alpha^2-2\alpha\varepsilon+\varepsilon^2  ,\hspace*{1cm} A_{2}=2\alpha^2+2\alpha\varepsilon+\varepsilon^2\nonumber , \hspace*{1cm} A_{3}=2\alpha^3-\alpha\varepsilon^2+\varepsilon^3 , \\
A_{4}&=&-2\alpha^3+\alpha\varepsilon^2+\varepsilon^3 ,\hspace*{1cm} A_{5}=2\alpha^3-\alpha\varepsilon^2-\varepsilon^3  .
\end{eqnarray}
What should be emphasized, although the nonlinearity constant $\chi$ does not appear here as a result of the fact that eigen-energies corresponding to the vacuum and one-photon states are equal to zero, the process of cutting of the  wave-function requires the presence of the nonlinearity.

\begin{figure}[h]
\begin{center}
\includegraphics[width=0.7\textwidth]{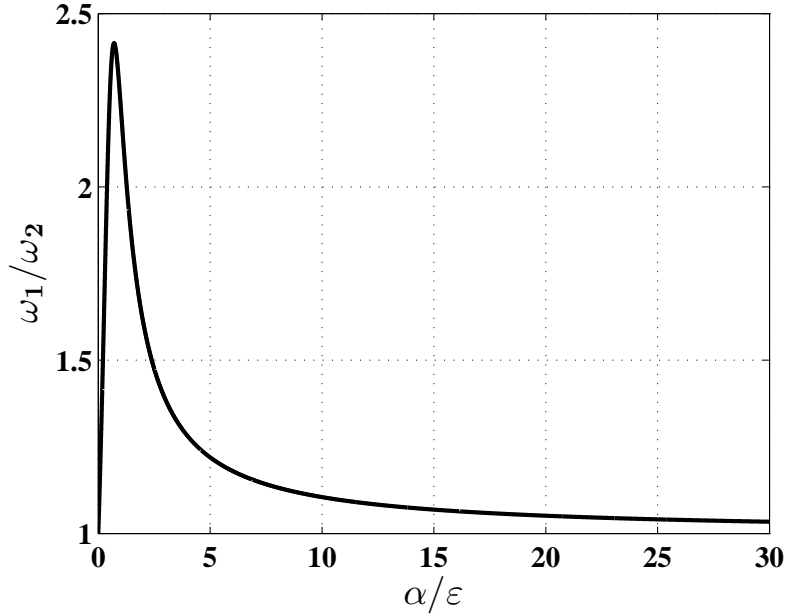}
\end{center}
\caption{The ratio $\omega_1/\omega_2$ as a function of  $\alpha/\varepsilon$ calculated from analytical solution (eq.\ref{eq9}).}
\label{omega1_omega2}
\label{fig2}
\end{figure}

We see that two frequencies $\omega_1$ and $\omega_2$ appear in our solution. To get regular and periodic solutions we have to chose the values of these frequencies carefully, and concentrate on the cases for which the ratio between them fulfil some special conditions. Thus Fig.\ref{omega1_omega2} shows that the ratio changes its value from $1$ to $\approx 2.4$. At this point the most promising values for  are $1$ and $2$. However, the cases when $\omega_1/\omega_2=1$ correspond to the situations when $\alpha =0$ (we have no external excitation) or $\varepsilon\rightarrow 0$ (internal interactions between oscillators are neglected), and are not interesting for us. Thus, the cases when $\omega_1/\omega_2=2$ seem to be more promising and therefore, we shall concentrate on them in this paper. From Fig.\ref{omega1_omega2} we see that assumed ratio between two frequencies can be achieved for two situations. Indeed, it can be shown that when the strengths of interactions $\alpha$ and $\varepsilon$ fulfil the following relation:
\begin{equation}
\varepsilon\,=\,\frac{12\,\alpha}{10\pm \sqrt{28}}
\label{2_eps}
\end{equation}
the condition $\omega_1/\omega_2 =2$ is achieved. For these two situations our system exhibits periodic behaviour with periods $T$ determined by the formula:
\begin{equation}
T=\frac{\sqrt{5\pm\sqrt{7}}\pi}{2\alpha},
\label{12}
\end{equation}
respectively.

\begin{figure}[h]
\begin{center}
\includegraphics[width=0.7\textwidth]{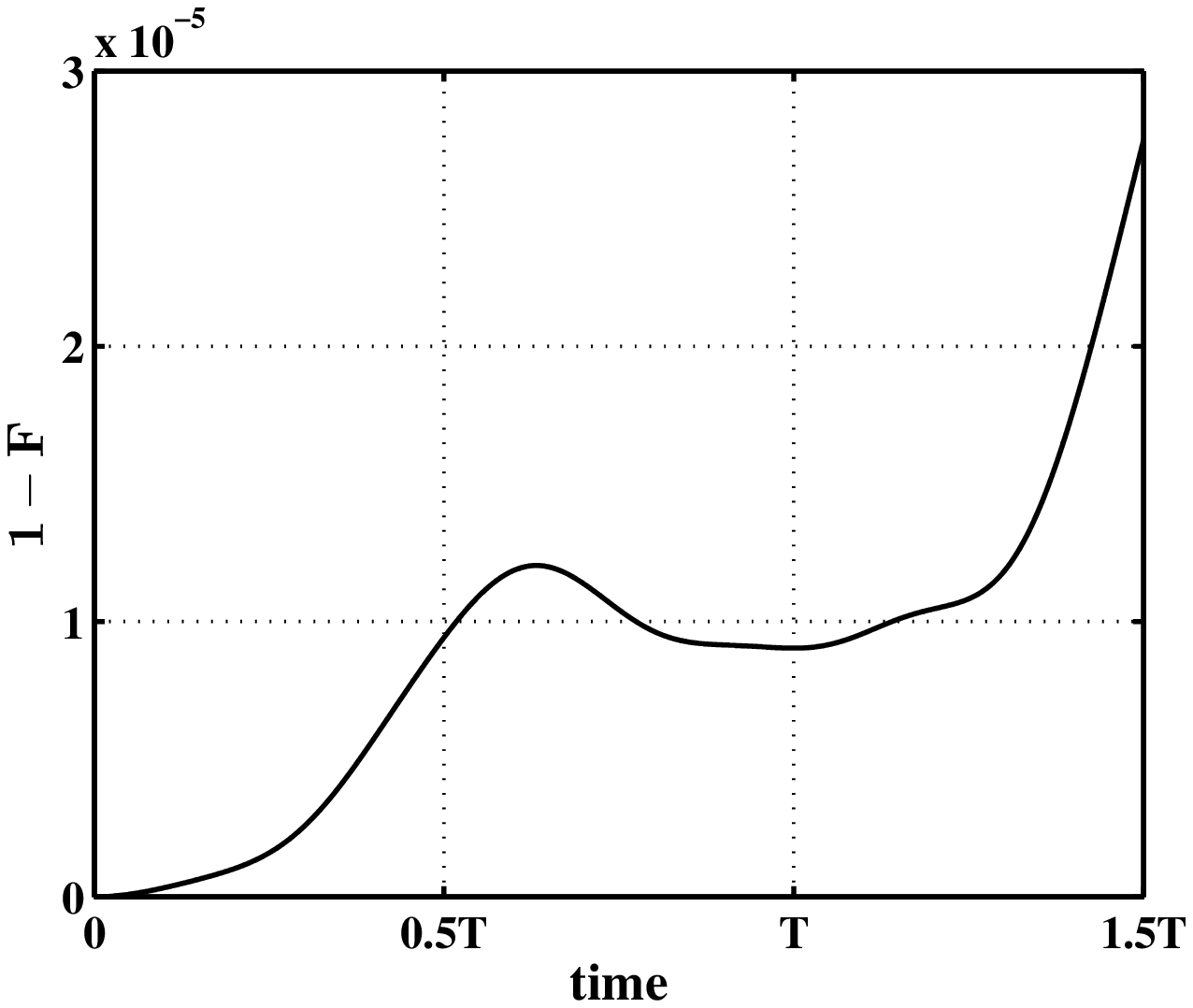}
 \end{center} 
\caption{Time-evolution of the parameter $1 - F(t)$ for $\alpha=0.001$, ~${\varepsilon}=\frac{12 \alpha}{10 + \sqrt{28}}$. 
Time is measured in units of $T=\frac{\sqrt{5+\sqrt{7}}\pi}{2\alpha}$.}
\label{fig3}
\end{figure}
To validate the exactness of NQS approximation we have solved Schr\"odinger equation numerically, assuming that $10$ $n$-photon states are involved in the evolution for each mode of the filed. Next, we have compared numerical results with those derived from our analytical formulas (valid for  $2\otimes 2\otimes 2$-dimensional Hilbert space). In particular, we have calculated the fidelity between the cut wave-function $|\psi\rangle_{cut}$ defined in eq.(\ref{eq:cut_wave_fun}) and ''full'' wave-function obtained from numerical analysis. Thus, Fig.\ref{fig3} shows the deviation of the fidelity $F(t)=\vert \langle\psi(t)\vert\psi(t)\rangle_{cut}\vert$ from the unity for one of the two cases in which we are interested in. For the second value of ${\varepsilon}$, $1-F(t)$ dependance is simmilar.  We see that such deviations are of the order of $\sim 10^{-5}$ 
and hence, we can assume that NQS approximation works well and gives sufficiently accurate results for our purposes.

\begin{figure}
\begin{center}
\includegraphics[width=0.9\textwidth]{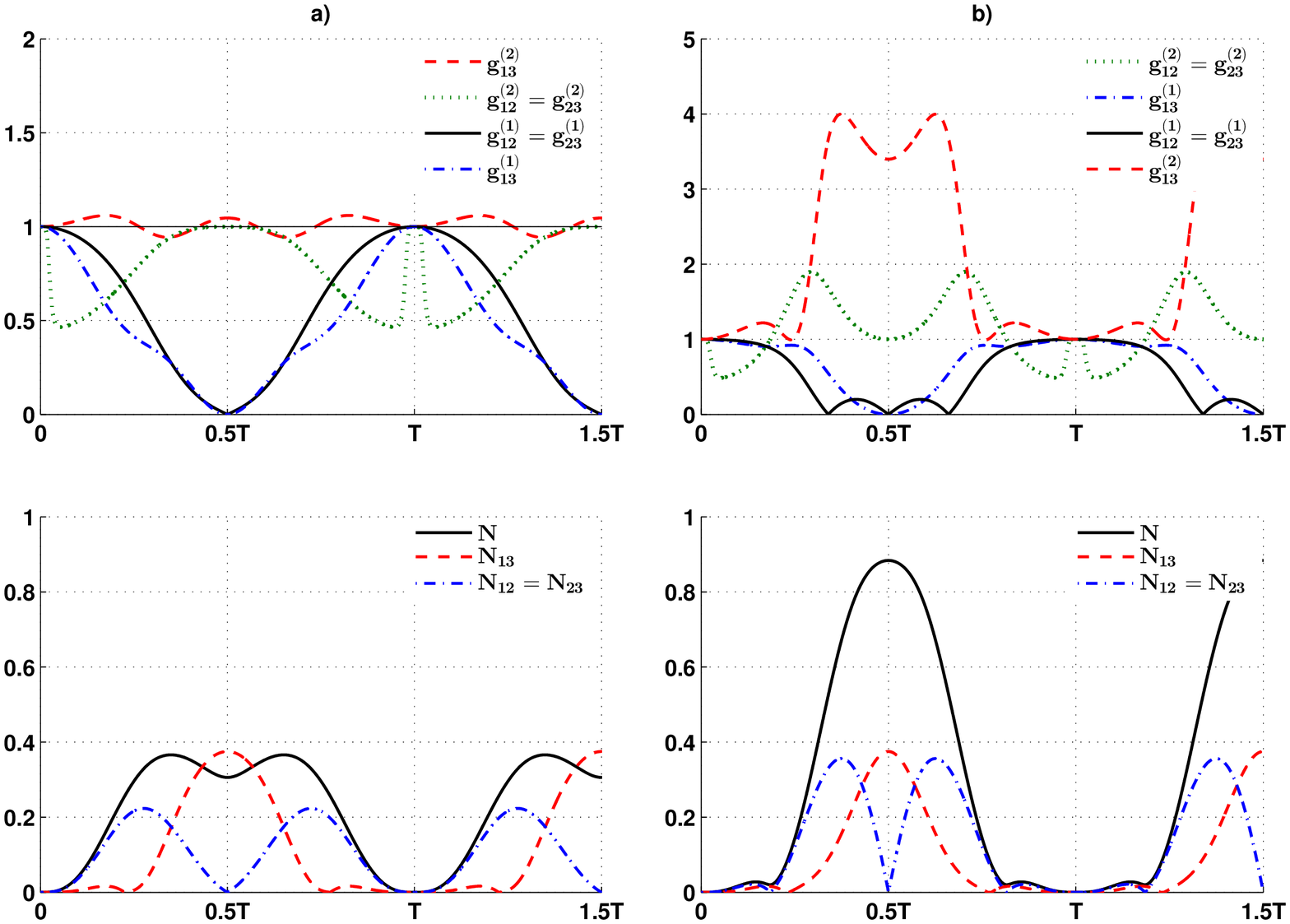}
\end{center}
\caption{Time evolution of:  a)~first-order and second order correlation function on the top, reduced (considered here) and tripartite (discussed in the next section) negativities  on the bottom for $\alpha=0.001\chi$, ${\varepsilon}=\frac{12 \alpha}{10 + \sqrt{28}}$, 
 b)~the same as in a) but for ${\varepsilon}=\frac{12 \alpha}{10 - \sqrt{28}}$. Time is measured in units of  $T$.}
\label{fig4}
\end{figure}

\section{First and second order correlation functions and tripartite entanglement}

The quantumness of the physical system can reveal for example in generation of various quantum correlations.
We will analyse quantum properties of the coupled nonlinear system using first and second-order correlation functions ($g^{(1)}$ and $g^{(2)}$), bi- and tripartite entanglement. 

Quantum correlations  and their relations to  entanglement creation  can be found for example in considering atomic ensemble in high-Q cavity \cite{SLG11}, optomechanical system composed of a cavity mode interacting with three bosonic modes obtained in an optical lattice \cite{SLF12}, or other optomechanical systems like cavity with oscillating mirror and filled with atoms \cite{IG08}, or nanomechanical resonator coupled to superconducting microwave cavity mode \cite{VT07}.  It was also suggested there that whenever the first order correlation (coherence) between the two considered modes is present, then there is no  entanglement between these modes.

The degree of coherence (amplitude correlation) can be defined with use of first-order correlation function~\cite{GK05, WM08}
\begin{equation}
g^{(1)}_{jk}=\frac{\vert \langle \hat{a}_{j}^{\dagger} \hat{a}_{k}\rangle \vert} {\langle \hat{a}_{j}^{\dagger}\hat{a}_{j}\rangle^{\frac{1}{2}} \langle \hat{a}_{k}^{\dagger}\hat{a}_{k}\rangle^{\frac{1}{2}}},
\label{g1}
\end{equation}
where $j, k$ label the system's modes and when $j\neq k$,  $g^{(1)}$ function describes cross-coherence between the two modes. The first-order correlation function takes values from zero to unity. If $g^{(1)}_{jk}=1$ one can observe coherence between the modes $j$ and $k$,  if it is zero - coherence is not present.
Second-order correlation (intensity correlations) is expressed by~\cite{GK05, WM08}:
\begin{equation}
g^{(2)}_{jk}=\frac{\langle \hat{a}_{j}^{\dagger}\hat{a}_{k}^{\dagger}\hat{a}_{j}\hat{a}_{k}\rangle} {\langle \hat{a}_{j}^{\dagger}\hat{a}_{j}\rangle\langle \hat{a}_{k}^{\dagger}\hat{a}_{k}\rangle}\, .
\label{g2}
\end{equation}
For correlated modes, function $g^{(2)}_{jk}$ takes values greater than unity; for uncorrelated modes is equal to unity, and it takes  values smaller than unity if the modes are anticorrelated. 
For the system of the three coupled nonlinear oscillators initially prepared in a vacuum state, and evolving according to  Schr\"odinger equation, the following relation between the amplitudes is satisfied: $C_{001}$ = $C_{100}$ and $C_{011}$ = $C_{110}$. Due to this symmetry, cross-coherence and second-order correlation function between boundary and central oscillators (b -- c) are equal to each other ($g_{12}^{(1)}=g_{23}^{(1)}$ and $g_{12}^{(2)}=g_{23}^{(2)}$). 

As a measure of 2-mode entanglement we apply that one which is  based on negative partial transposition criterion (NPT criterion) -- the negativity \cite{P96,HHH96}. If $\rho_{ij}^{T_i}$ describes the partial transposition (with respect to $i$ mode) of the density matrix for the system, then the negativity is defined as a sum of absolute values of all the negative eigenvalues calculated for $\rho_{ij}^{T_i}$:
\begin{equation}
N_{ij}(\rho)=\sum\limits_{l} \mu_{l}(\rho_{ij}^{T_i})\, ,
\label{neg_ij}
\end{equation}
where $\rho_{ij}=Tr_k(\rho_{ijk})$. Negativity is able to distinguish between separable and entangled states for the systems $2\otimes 2$ or $2\otimes 3$ and  takes values between 0 -- for separable states  and 1 -- for maximally entangled ones for such systems. 
The evolution of all the discussed types of quantum correlations for bipartite systems is presented in Fig.\ref{fig4}  for the two previously chosen values of $\varepsilon$, see eq.(\ref{2_eps}). For both values of coupling $\varepsilon$, the bipartite entanglement is created alternately for b -- c oscillators and b -- b ones. Therefore the boundary modes, even though not coupled to each other, can create entangled states via the interactions with the central  oscillator. 
The degree of entanglement between boundary oscillators is not sensitive to the coupling strength $\varepsilon$. 
On the contrary, the degree of entanglement created by b -- c oscillators is directly dependent on $\varepsilon$. By changing the value of internal interactions one can influence the degree of neighbouring bipartite entanglement, and additionally the degree of tripartite entanglement (discussed later).

It is known that intermode correlations of the 2$^{nd}$ order are related to the process of bipartite entangled states creation. 
In the chain of nonlinear oscillators, 2$^{nd}$ order correlations in boundary modes are connected to entanglement between them if these modes are not additionally correlated (or anticorrelated) with the mode of central oscillator. On the other hand entanglement between the modes of b --- c oscillators can arise if they are anticorrelated and no correlations between the boundary oscillators are present. What is also worth sterssing is the relation between the 1$^{nd}$ order  correlation functions and bipartite entanglement mentioned in \cite{SLF12}. In the chain of 2-qubit systems 1$^{nd}$ order coherence is connected with lack of entanglement when the systems periodically returns to its initial vacuum state (for times distant by one period - T value). During the exchange of entanglement between the pairs of oscillators one can find that the maximum entanglement between the b --- b modes (not directly interacting with each other) is created when there is no coherence between them. The opposite is observed for interacting modes.

The whole system is composed of three oscillators (three modes of the field), and as it was already shown, each of them can evolve as a two-level system. In consequence, each oscillator can be treated as a qubit, and the whole system as a 3-qubit one.  In this section we will analyse our system's ability to produce tripartite entanglement and identify different classes of 3-qubit entangled states.

Whereas the entanglement of bipartite systems is well understood, it was soon realized that the entanglement of tripartite quantum states is not a trivial extension of its bipartite counterpart \cite{HHH09}.

D\"ur, Vidal and Cirac \cite{DVC00} have classified three-qubit states according to their equivalence types under stochastic local operations and classical communication (SLOCC). Thus, they have proposed three classes of the three-qubit states: full separable states (labeled here as I), biseparable states (II) and full tripartite entanglement states (III). From the other hand, for the pure three-qubit states, two inequivalent kinds of tripartite entanglement are distinguished, represented by the GHZ and W-states \cite{DVC00}. The GHZ state has only full tripartite entanglement and the entanglement disappears if any one of the three qubits is traced out -- the remaining two are always in a separable state. On the other hand, for the W-states bipartite entanglement is present, and when one of the qubits is removed by the tracing procedure, the other two remain maximally entangled. 
However, one should keep in mind that classification proposed in \cite{DVC00} does not include some other states, such as the so called star-shaped states \cite{PB03a,PB03b}. Therefore, Sab\'in and G. Garc\'ia-Alcaine have proposed in  \cite{SG08} an extended classification that include both pure and mixed states. To be more specific, classification scheme proposed in  \cite{SG08}, divides the class of tripartite entangled states on the four following subtypes:
\begin{itemize}
\item subtype III-0 that contains the states for which the all bipartite entanglements  disappear (all reduced negativities $N_{ij}\neq 0$ -- see eq.(\ref{neg_ij})). The well-known representative state belonging to this subtype is GHZ state. From that reason the states from this class are called GHZ-like states;
\item subtype III-1 for which only one of the reduced negativities is non-zero, whereas remaining two are zero valued;
\item subtype III-2 contains star-shaped states for which two reduced negativities have non-zero values;
\item subtype III-3. To this subtype belong so called W-like states characterized by three non-zero reduced negativities.
\end{itemize} 

To sum up the above description, all possible classes of the tripartite entanglement and their relation with the reduced bipartite negativities has been presented in Tab.~\ref{tab1}.
It is worth emphasizing that D\"ur, G. Vidal and J. Cirac classification scheme contains states of the two types only. There are: types III-0 and III-3.
\begin{table}[h]
\begin{center}
\begin{tabular}{|c|c|c|}
\hline
type of tripartite& reduced& generated \\
entanglement & entanglement&  state  \\
\hline
III-0 & $N_{ij}=N_{jk}=N_{ik}=0$ &GHZ-class state\\
\hline
III-1 & $N_{ij}\neq 0,\;N_{jk}=N_{ik}=0$ &\\
\hline
III-2 & $N_{ij}\neq 0,\;N_{jk}\neq 0,\;N_{ik}=0$ &star-like state\\
\hline
III-3 & $N_{ij}\neq 0,\;N_{jk}\neq 0,\;N_{ik}\neq0$ &W-class state\\
\hline
\end{tabular}
\caption{The types of full tripartite entanglement.\label{tab1}}
\end{center}
\end{table}

Now, in order to classify various types of the entanglement which can be generated by the system, we follow the path proposed in \cite{SG08} applying geometrical average of three negativities:
\begin{equation}
N(\rho)=\left( N_{1-23} N_{2-13}N_{3-12}\right)^{\frac{1}{3}},
\label{neg}
\end{equation}
as a good measure of full tripartite entanglement. The parameters $N_{i-jk}$ are the bipartite negativities for three-mode density matrix $\rho_{ijk}$, where the partial transpose is made for the mode (subsystem) $i$. By contrast to other tripartite entanglement measures (see for instance \cite{YS04,MCF14}), the one defined in (\ref{neg}) is always equal to zero for the states belonging to the classes I and II (according to the classification proposed by D\"ur, Vidal and Cirac), and has non-zero values  for all subtypes of tripartite entanglement listed above.

Time evolution of negativities describing tripartite entanglement for the two chosen coupling strengths is presented in Fig.\ref{fig4}. It is seen that for both values of $\varepsilon$, 3-qubit entangled states are generated. We have already shown that coupling strength influences the degree of 2-qubit entanglement between b -- c oscillators, but we may also conclude that this influence on the formation of 3-qubit entanglement is more pronounced. Even though Bell-like states are obtained with probabilities that do not exceed $40$\%, the 3-qubit entangled states are obtained (for the same values of the interaction strengths) with probabilities higher than $80$\% -- Fig.\ref{fig4}. 

\begin{figure}
\raisebox{7cm}{\textbf{a)}}
\includegraphics[width=0.6\textwidth]{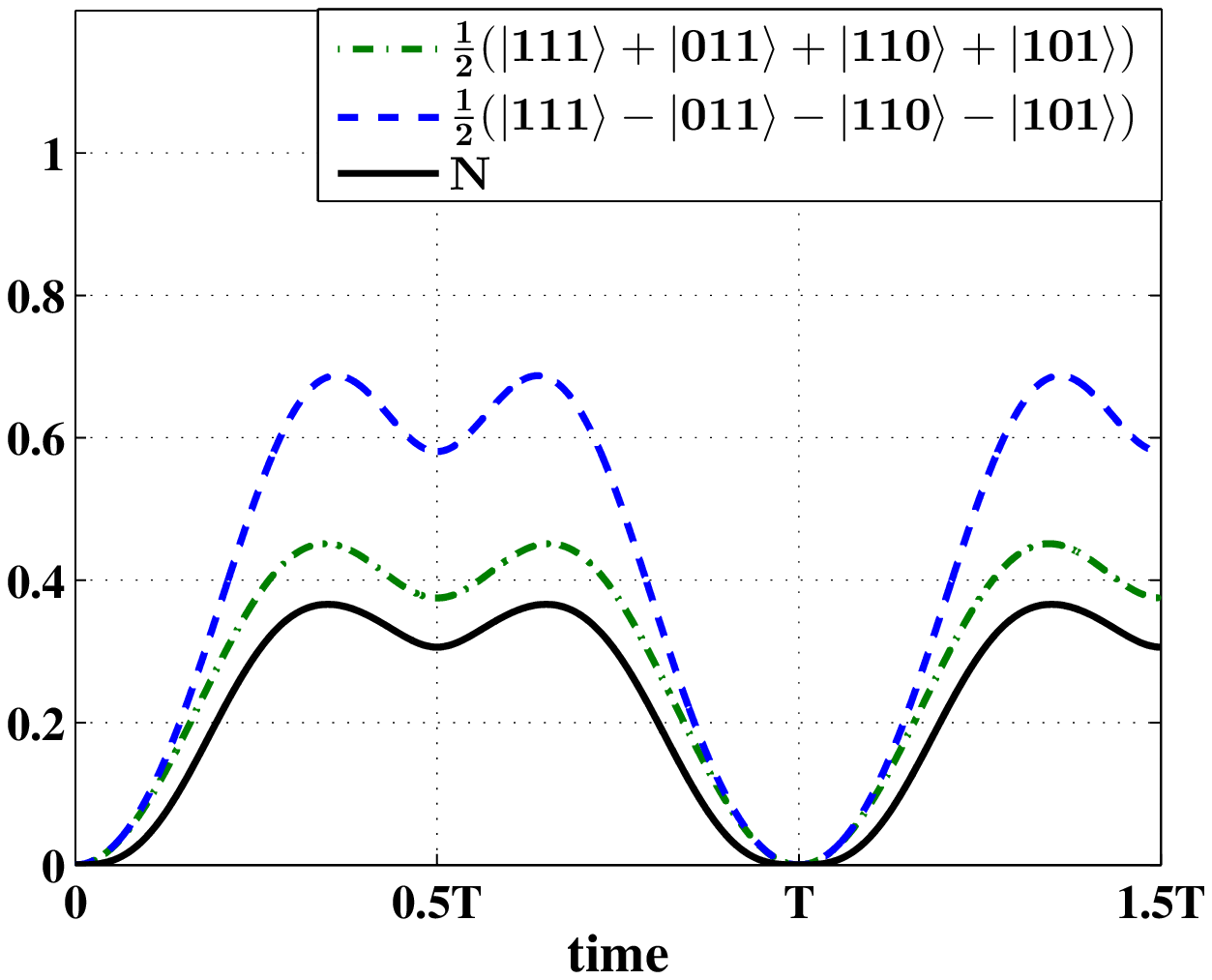}\\
\raisebox{7cm}{\textbf{b)}}
\includegraphics[width=0.6\textwidth]{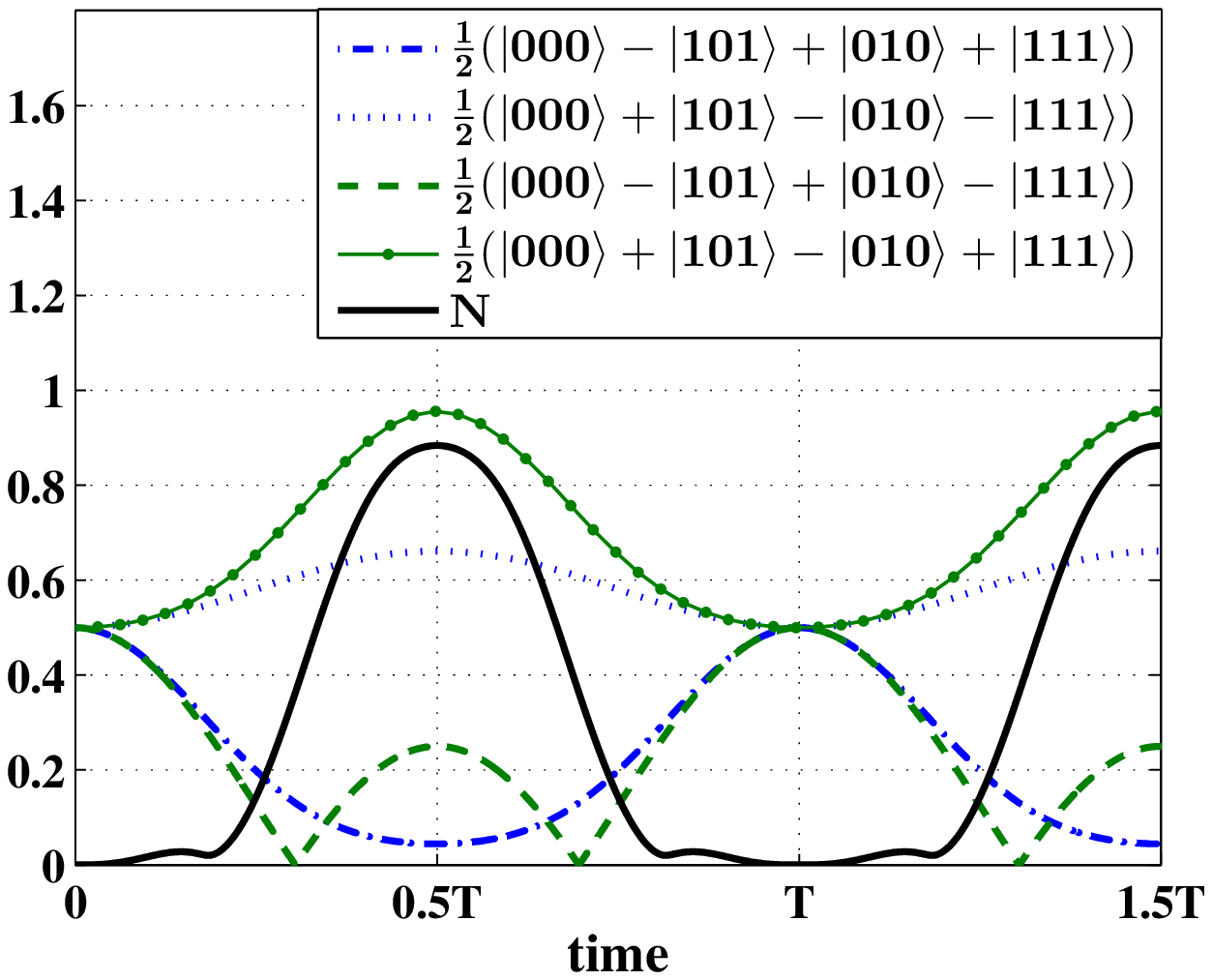}
\caption{Time evolution of  fidelities corresponding to same states and tripartite negativity for $\alpha=0.001\chi$ for: a)~ ${\varepsilon}=\frac{12 \alpha}{10 + \sqrt{28}}$, $T=\frac{\sqrt{5+\sqrt{7}}\pi}{2\alpha}$ and b)~${\varepsilon}=\frac{12 \alpha}{10 - \sqrt{28}}$. Time is measured in units of $T=\frac{\sqrt{5-\sqrt{7}}\pi}{2\alpha}$.}
\label{fig5}
\end{figure}

For both $\varepsilon$ values, during the evolution in time, different types of tripartite entangled states  can be obtained. For times equal to $0.5$ of the oscillation period ($t=0.5T$) we have identified the entangled state of the whole system as being the  III-1 type (Tab.\ref{tab1})
 -- for that  moments of time only boundary oscillators are entangled, but the system as a whole can be found in almost maximally entangled state Fig.\ref{fig4}. The b --- b oscillator pair is correlated and both of the b --- c oscillators pairs are not correlated. Entangled W-state is generated with smaller probability; as seen from Fig.\ref{fig4} all of the 2-qubit pairs are entangled  and all of them are correlated ($g^{(2)}_{12,23}>0$ and $g^{(2)}_{13}>0$).
 
 Smaller interaction between boundary and central oscillators result in smaller probabilities of the obtained tripartite entangled state --- Fig.4. For other times, for which one can observe entanglement between all of the oscillators, the system can be found in III-3 type states -- W-states. Adequate fidelities corresponding to different 3-qubit entangled states are presented in Fig.~\ref{fig5}.

For smaller b - c interaction, 
when tripartite negativity reaches its maximum, all of the three reduced negativities are non-zero (III-3 type of tripartite entanglement). We can conclude that this class of tripartite entanglement is associated with production of W-class state  --- Fig.\ref{fig5}a. As seen from Fig.\ref{fig4} the W-class state is obtained when all oscillators pairs are anticorrelated. Additionally, generation of III-1 type states is also connected with second order correlations present only in the modes corresponding to boundary oscillators.
For larger interaction between the oscillators (Fig.~\ref{fig5}b), maximum of tripartite entanglement is associated with production of the states $(\vert 0 \rangle_{2}\pm\vert 1 \rangle_{2})(\vert 00\rangle_{13}\pm\vert 11\rangle_{13})$. 

If compare our results with the ones presented in \cite{SWU06} for the system of three mutually coupled Kerr-like oscillators with no external excitation, for which the authors reported the possibility of producing entangled W-states, 
we can see that the dynamics in our model is richer and the system is able to produce other types of 3-qubit entangled states with high probabilities.

\begin{figure}
\includegraphics[width=0.6\textwidth]{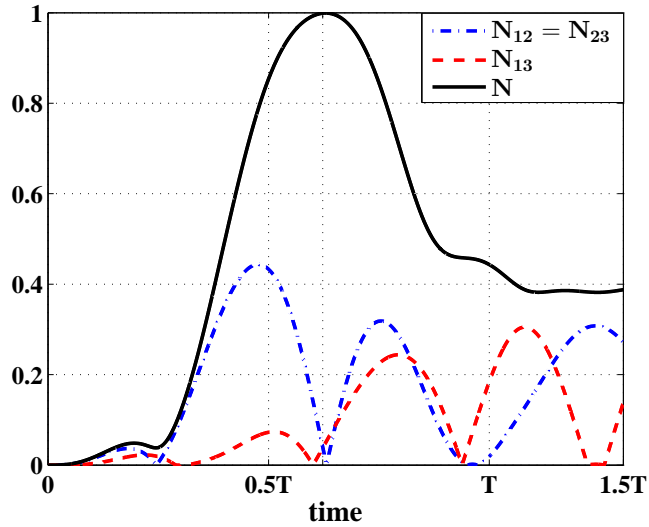}
\caption{Time evolution of reduced and tripartite negativities for ${\varepsilon}=\frac{12 \alpha}{10 - \sqrt{28}}+\Delta$, $\Delta=-0.6\alpha$. Time is measured in units of the period $T$, used in Fig.\ref{fig5}b.}
\label{fig6}
\end{figure}

The results presented so far concern the two values of interactions between the neighbouring oscillators that correspond to regular and periodic solutions. If the values of $\varepsilon$ are different, evolution of correlation functions and entanglement will be more complicated. But it appears that it is possible to obtain values of negativity describing tripartite entanglement slightly less than unity -- meaning that almost maximally entangled 3-qubit state (MES) is generated. That can be obtained by slightly detuning the value of $\varepsilon$ from one of the regular solutions. The results are seen in Fig.\ref{fig6} that presents all of the 2-qubit negativities and a 3-qubit one (note that the time is scaled in the same units of $T$, as in Fig.\ref{fig5}b for periodic behaviour). The states that are generated  are:  $\frac{1}{\sqrt{2}}(\vert 000\rangle\pm\vert 111\rangle)$ and $\frac{1}{\sqrt{2}}(\vert 010\rangle\pm\vert 101\rangle)$.
Unfortunately, they are slightly perturbed by states $\vert 001\rangle$ and $\vert 100\rangle$ and the result is that we cannot  produce maximally entangled GHZ state (III-0 in Tab.I) with probability 1.  Additionally,  once this state is generated, the 2-qubit entangled states are produced simultaneously for boundary and between boundary and central oscillators. Due to the fact that they oscillate with slightly different frequencies,  in time we will observe previously mentioned flow of entanglement between the oscillator's pairs and production of other types of 3-qubit entangled states.

\section{Damped system}
\subsection{Amplitude damping}
\label{ampl_damp}

In this section we shall discuss how damping processes influences our system's dynamics. In particular, the time-evolution of the first- and second order correlation functions as well as  bi- and tripartite entanglement will be considered.

When amplitude damping is assumed (for the zero-temperature bath) the system's evolution is governed by the following master equation~\cite{WM08, GZ00}: 
\begin{eqnarray}
	\label{mastereq1}
\frac{d\hat{\rho}}{dt}&=&-\frac{1}{i}\left(\hat{\rho}\hat{H}-\hat{H}\hat{\rho}\right) \\
& &+\sum\limits_{j=1}^{3} \frac{\kappa_{j}}{2}\left( 2 \hat{a}_{j} \hat{\rho} \hat{a}^\dagger_{j} - \hat{a}^\dagger_{j} \hat{a}_{j} \hat{\rho} - \hat{\rho} \hat{a}^\dagger_{j} \hat{a}_{j} \right), \nonumber
\end{eqnarray}
where we have introduced damping parameter  $\kappa_j$ ($j=1,2,3$) describing interaction with zero temperature bath for the modes $1$, $2$, $3$, respectively. For simplicity, we assume that all damping parameters appearing here are the same for all the modes \textit{i.e.} $\kappa_1=\kappa_2=\kappa$ and  $\kappa=0.1\alpha$. The same as for the cases discussed in previous sections, the system is excited in two modes and $\alpha=0.001\chi$ (only boundary oscillators $1$ and $2$ are excited). Moreover, since the results corresponding to the two considered values of the excitation strength $\varepsilon$ are very similar to each other, we shall concentrate on the case when ${\varepsilon}=\frac{12 \alpha}{10 - \sqrt{28}}$ (stronger internal interaction case). 
We shall mostly discuss two cases: one corresponding to the periodic solution, \textit{i.e.} for $\varepsilon  = 12\alpha /(10-\sqrt{28})$ and the second which corresponds to the situation presented in Fig.\ref{fig6}, when $\varepsilon  = 12\alpha /(10-\sqrt{28})+\Delta$ and the system exhibits quasi-periodic behaviour.

\begin{figure}[h]
\begin{center}
\raisebox{7cm}{\textbf{a)}}
\includegraphics[width=0.625\textwidth]{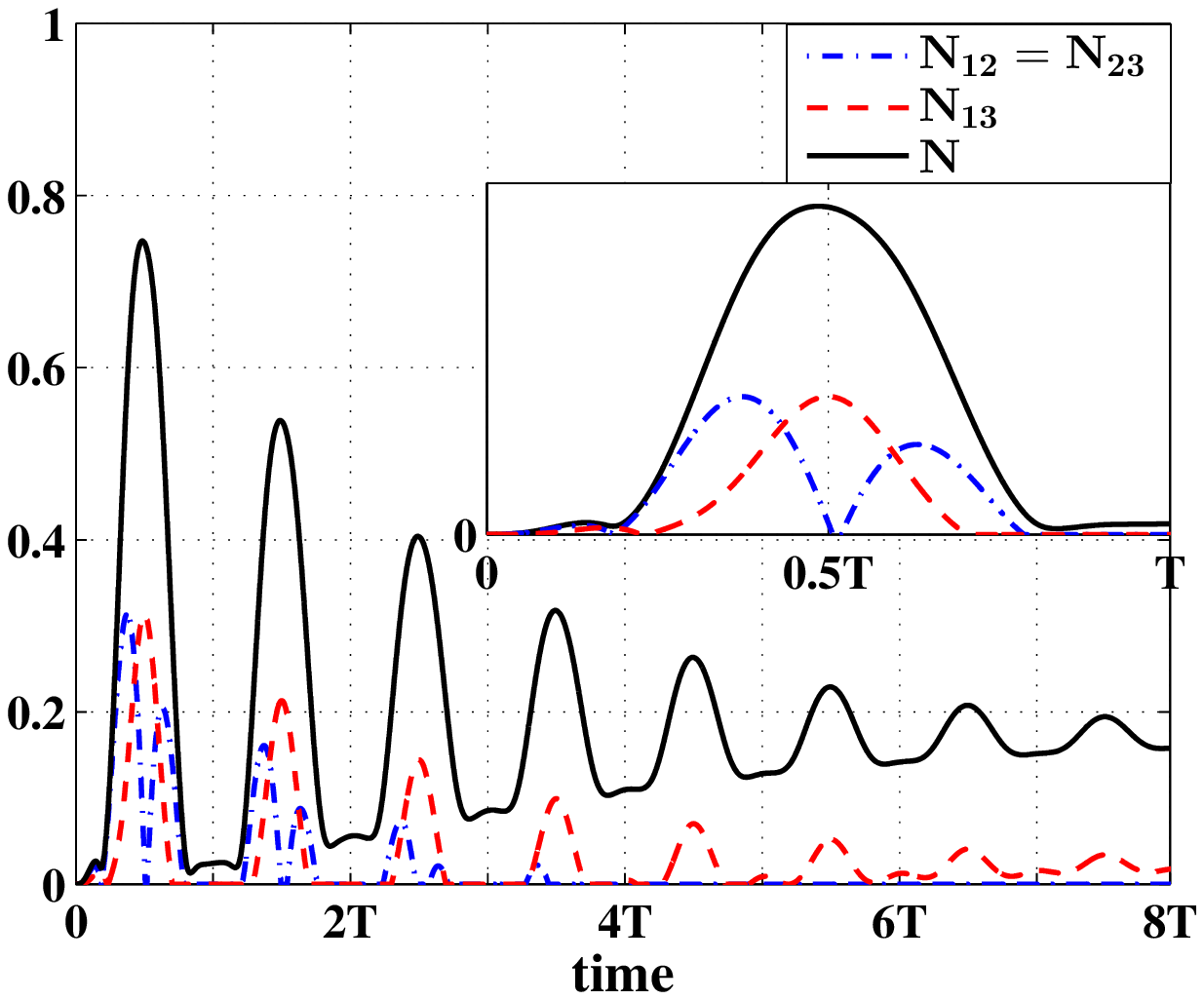}\\
\raisebox{7cm}{\textbf{b)}}
\includegraphics[width=0.625\textwidth]{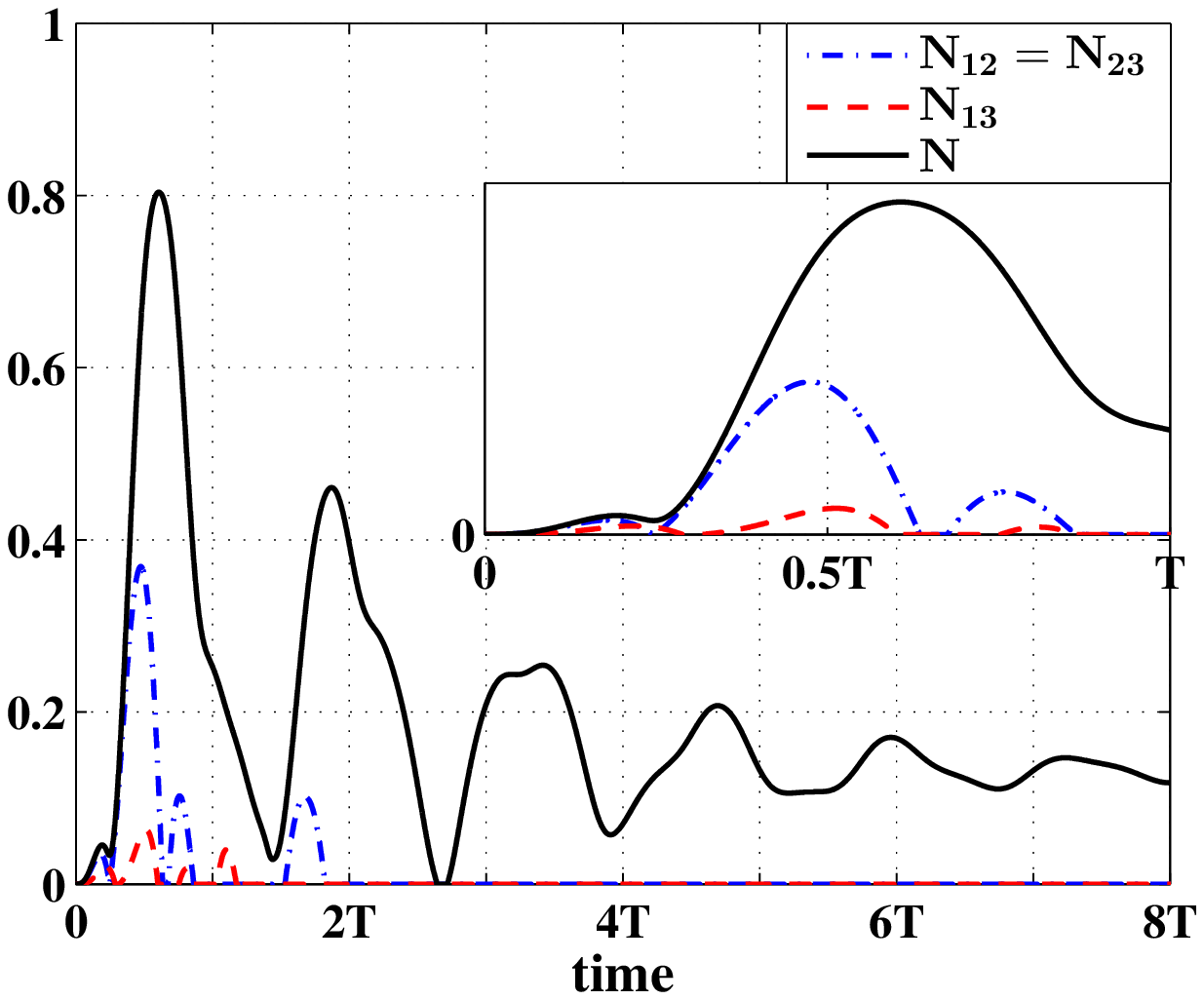}
\end{center}
\caption{Time evolution of the reduced and tripartite negativities for $\alpha=0.001\chi$, ${\varepsilon}=\frac{12 \alpha}{10 - \sqrt{28}}$ (a), ${\varepsilon} =\frac{12 \alpha}{10 - \sqrt{28}}+\Delta$, where $\Delta =-0.6\alpha$ (b). The amplitude damping parameter $\kappa=0.1\alpha$. Time is measured in units of  $T=\frac{\sqrt{5-\sqrt{7}}\pi}{2\alpha}$.}
\label{fig7}
\end{figure}

When our system is influenced by an external zero-temperature bath, its behaviour changes considerably from that corresponding to the non-damped cases. In Fig.~\ref{fig7} we show the time-evolution of the negativities describing the both: bi- and tripartite entanglement ($\{N_{12}=N_{23}, N_{13}\}$ and $N$, respectively). Fig.\ref{fig7}a corresponds to he case when $\omega_1=\omega_2$, \textit{i.e.} the system evolves periodically, whereas Fig.\ref{fig7}b shows system's evolution when internal coupling constant $\varepsilon$ is slightly perturbed by $\Delta=-0.6\alpha$. (for such  situation we get maximal value for the tripartie negativity for not damped case -- see Fig.\ref{fig6}). Moreover, when internal coupling constant is perturbed, our system exhibits quasi-periodic evolution. Thus,
we see that $N$ exhibit damped oscillations and tends to its final non-zero value ($\sim 0.16$). It differs from zero, so tripartite entanglement does not disappear during the whole time of the system's evolution. Moreover, the maximal value of $N$ is reached for the time $t=T/2$ and its value $\sim 0.7$. At the same moment of time also $N_{13}$ becomes maximal ($\sim 0.3$) whereas $N_{12}=N_{13}$ are equal to zero. In consequence, we do not observe bipartite entanglement between neighbouring oscillators whereas the entanglement between two boundary oscillators reaches its maximum (one should remember that it does not mean that we generate here maximally entangled state). For such a situation, the state of the III-1 type is generated. It is the state $\vert 0 \rangle_{2}\pm\vert 1 \rangle_{2})(\vert 00\rangle_{13}\pm\vert 11\rangle_{13})$ with some addition of the states: $\vert 001\rangle$, $\vert 100\rangle$, $\vert 011\rangle$, $\vert 110\rangle$ and hence, maximal value of $N<1$. Moreover, we observe sudden death of entanglement \cite{ZHH01,*YE04,*ILZ07,*BMW11} and its rebirth \cite{FT06,*BF06,*FT08} for all cases of bipartite entanglement represented by the negativities $N_{ij}$ ($i=\{1,2\}$, $j\{2,3\}$). In long-time limit we see that final values of the negativities $N$ and $N_{13}$ reach some non-zero values whereas the entanglement represented by $N_{1,2}=N_{23}$ disappear. 
When internal coupling is equal to $\frac{12 \alpha}{10 - \sqrt{28}}+\Delta$, where $\Delta =-0.6\alpha$ (Fig.\ref{fig7}b) we see  behaviour similar to that observed for the situation presented in Fig.\ref{fig7}a.  For this case we observe slightly greater value of the maximum of $N\simeq 0.8$ and damped oscillations of all negativities. The same as previously, sudden death of bipartite entanglement and its rebirth represented by all reduced negativities are present in the system's evolution, and $N$ tends to its final non-zero value again. However, due to the fact that two slightly different frequencies govern time-evolution of the model, some dephasing appear in the system, and when $N$ reaches its first maximum, all the reduced negativities become practically equal to zero. In consequence, this maximum of $N$ corresponds to the generation of the tripartite entangled state of the type III-0, contrary to the situation depicted in Fig.\ref{fig7}a, where the entanglement of type III-1 was present. Moreover, we see that for time $t\rightarrow\infty$ all  reduced negativities describing bipartite entanglement tend to zero.

\begin{figure}
\begin{center}
\raisebox{7cm}{\textbf{a)}}
\includegraphics[width=0.6\textwidth]{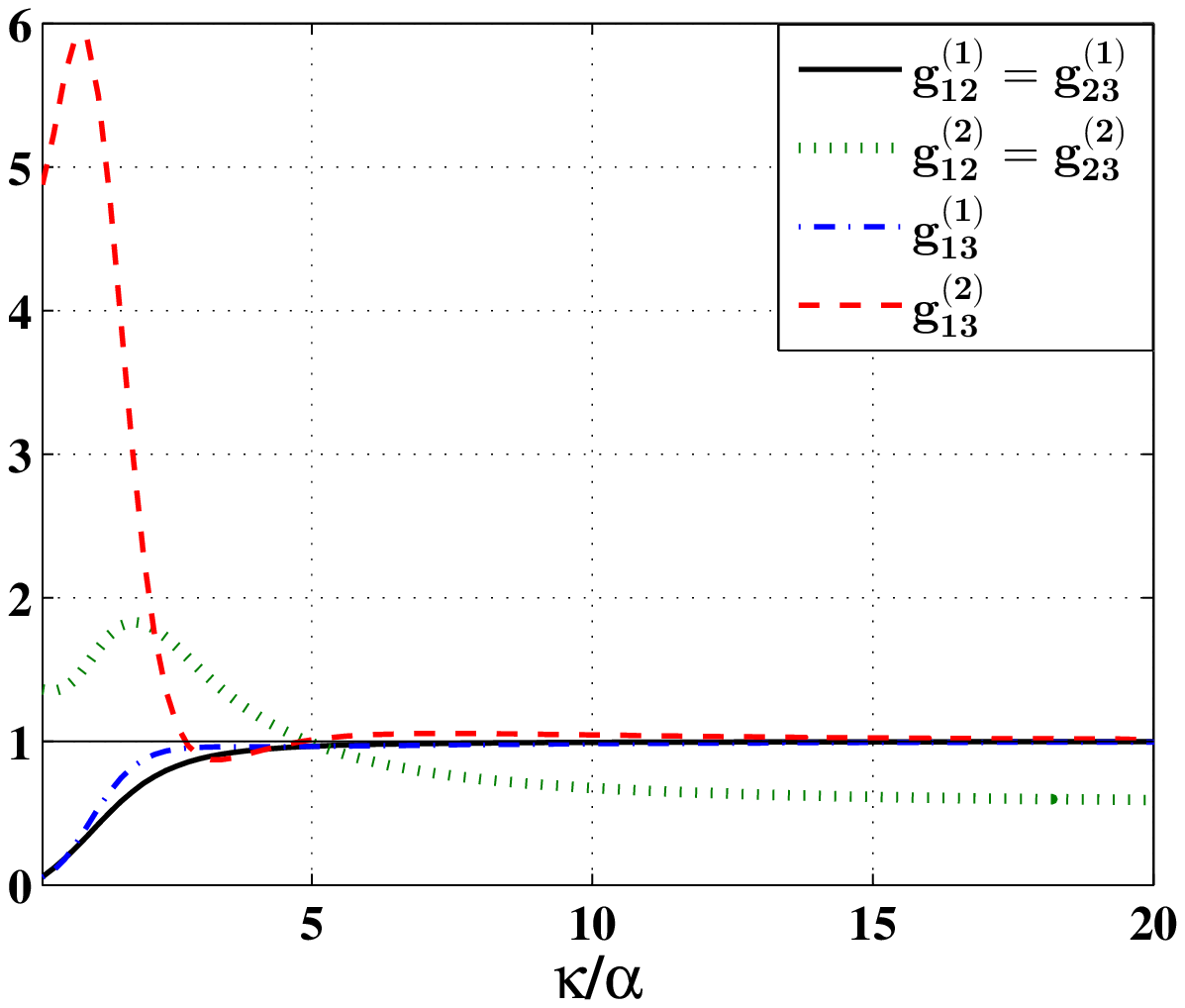}\\
\raisebox{7cm}{\textbf{b)}}
\includegraphics[width=0.6\textwidth]{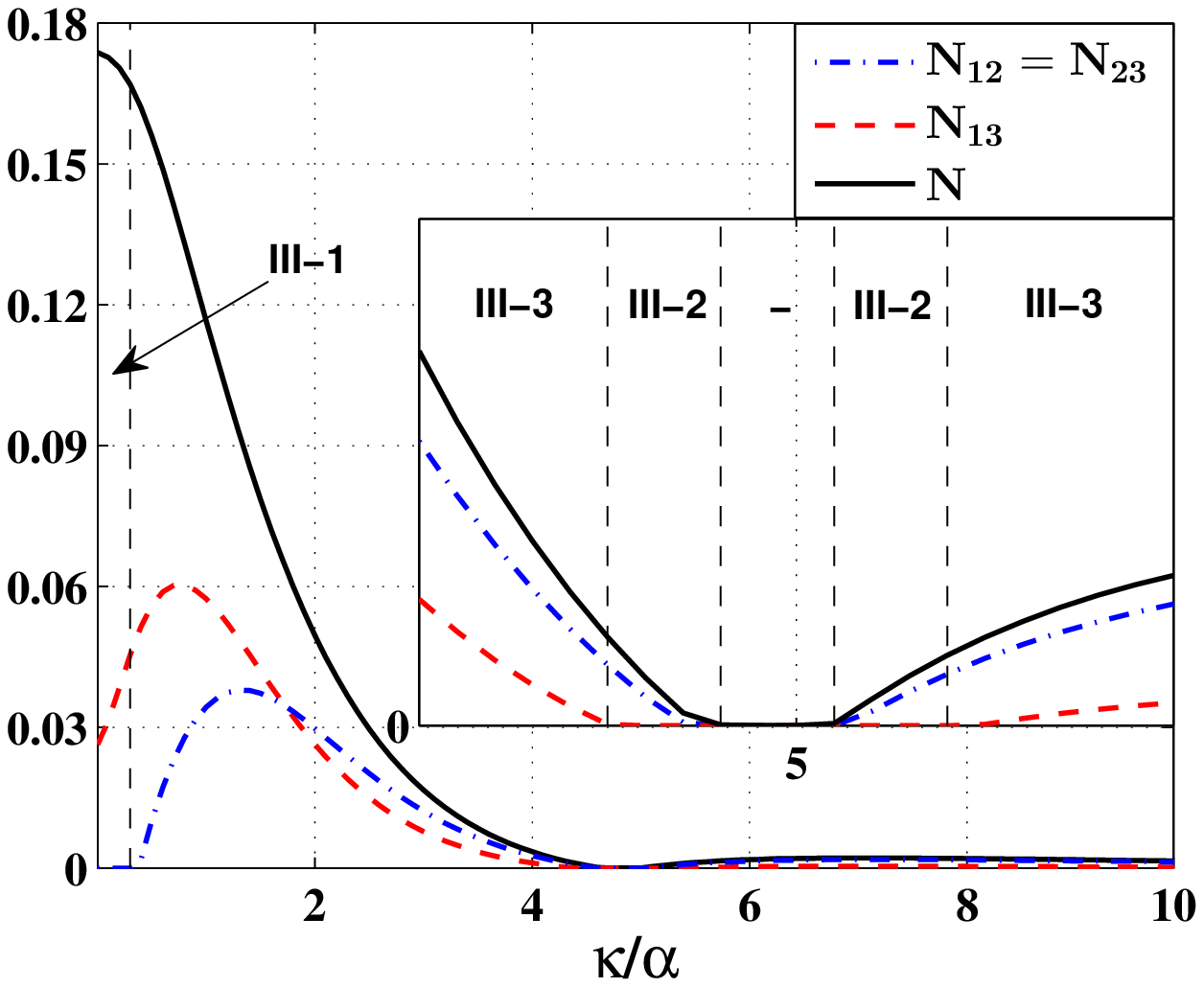}\\
\raisebox{7cm}{\textbf{c)}}
\includegraphics[width=0.6\textwidth]{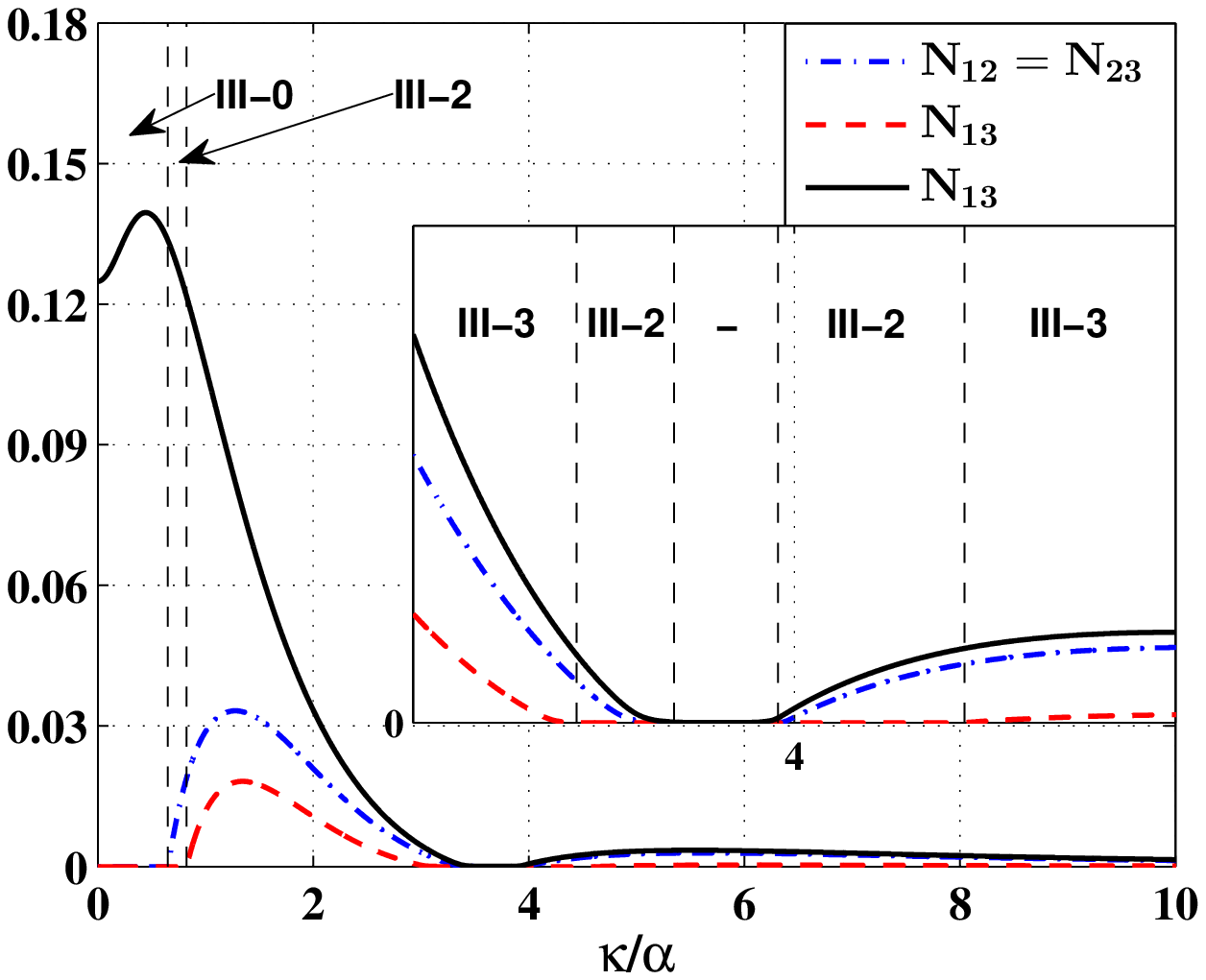}
\end{center}
\caption{Steady-state solutions for the the intermode correlation functions $g^{(1)}$ and $g^{(2)}$ (a), and for the negativities (full tripartite $N$ and reduced bipartite $N_{ij}$)  vs. value of the damping parameter where  ${\varepsilon}=\frac{12 \alpha}{10 - \sqrt{28}}$ (a,b) and  ${\varepsilon}=\frac{12 \alpha}{10 - \sqrt{28}}+\Delta$ ($\Delta =-0.6\alpha$) (c). The external coupling strength $\alpha=0.001\chi$.}
\label{fig8}
\end{figure}

Quantum correlations and hence, characteristics of our system's steady-state strongly depend on the value of the damping constant. Thus, Fig.~\ref{fig8} shows the dependence of all considered here negativities and correlation functions on the value of $\kappa$ (expressed in the units of external coupling $\alpha$).
Fig.~\ref{fig8}a depicts how first- and second-order intermode correlation functions change their final values with increasing value of $\kappa$ (for the periodic and quasi-periodic cases those dependences are almost identical). We see that for very weak damping case all $g^{(1)}$ functions are close to zero, so we do not observe first-order correlations. 
However, as 
\noindent$\kappa$ increases, when $\kappa /\alpha>\sim 5$, all functions practically become equal
to unity. If we look at second-order correlations, situation is more complicated. For small values of $\kappa$, they are greater than $1$ and we observe intermode anti-correlations for all modes of the field. Those anti-correlations increase to reach their maximal values and then, start to fall down with increasing $\kappa$. For the correlations between two boundary oscillators' modes ($1$ and $3$)  some range of the values of $\kappa$ exists for which anti-correlations are observed. However, for $\kappa >\sim 5$ we observe tiny and vanishing correlations again. Correlations between two neighbouring modes ($1-2$ and $2-3$) reaches their maximum for greater value of $\kappa$ if we compare it with that corresponding to the maximal value of $g^{(2)}_{13}$. Then, the values of $g^{(2)}_{12}$ and $g^{(2)}_{23}$ decrease and become negative for the same value of $\kappa$ when anti-correlations between the modes $1-3$ disappear. What is interesting, although the correlations described by $g^{(2)}_{13}$ practically disappear for stronger damping, anti-correlations described by $g^{(2)}_{12}$ and $g^{(2)}_{23}$ remains in the system even for large values of $\kappa$.

Figs \ref{fig8}b and \ref{fig8}c show long-time values of the tri- and bipartite  negativities, corresponding to the various   values of damping constant $\kappa$ (expressed in the units of external coupling strength $\alpha$). Fig.\ref{fig8}b corresponds to the case when system's dynamics is periodic (${\varepsilon}=\frac{12 \alpha}{10 - \sqrt{28}}$), whereas for the situation presented in Fig.\ref{fig8}c internal coupling strength is  perturbed \textit{i.e.}  ${\varepsilon}=\frac{12 \alpha}{10 - \sqrt{28}}+\Delta$, where $\Delta =-0.6\alpha$ (for this case our system exhibits quasi-periodic behaviour).
In Fig.\ref{fig8}b we see that, although the tripartite entanglement decreases with growing value of $\kappa$ (when $\kappa /\alpha <\sim 4.5$), the negativities describing bipartite entanglement can increase for such situation, and for some values of the damping constant they reach their maximal values. When 
$\kappa /\alpha <\sim 5.4$ (\textit{see} Tab.\ref{tab2} and inset in Fig.\ref{fig8}b) three various classes of the tripartite entangled states can be generated depending on the value of $\kappa$, and for $\kappa /\alpha\simeq5$ no entanglement is produced -- all the negativities are equal to zero.
In general, if we look at the values of the negativities describing two- ($N_{12}=N_{23}$, $N_{13}$) and tripartite ($N$) entanglement, we can identify various regions for which one or more of them are equal to zero. Thus, we can identify various classes of tripartite entangled states corresponding to different values of $\kappa$ (Tab.\ref{tab2} shows approximate values of $\kappa$ for which the steady-states of our system belong to various classes of the tripartite entangled state). In particular, for the situation presented in fig.\ref{fig8}b, they are III-1, III-2 (star-like) and III-3 (W class) states. It is seen from Fig.\ref{fig8}b that highest values of all bipartite negativities we can find when $\sim 0.5\kappa /\alpha<\sim 2$. For such   a situation tripartite negativity $N$ is relatively high as well, so W state can be generated with better efficiency then for the cases when other classes of states are produced. Nevertheless, applying various strengths of damping we can switch final system's state from one belonging to the particular class to another. 

Fig.\ref{fig8}c corresponds to the situation when internal coupling $\varepsilon$ is perturbed, \textit{i.e.} ${\varepsilon}=\frac{12 \alpha}{10 - \sqrt{28}}+\Delta$ ($\Delta =-0.6\alpha$). For such situation the system evolves quasi-periodically and hence, we observe some dephasing  effect as a result of the appearance of two different frequencies in the system's dynamics. In consequence, different types of the bipartite entanglement represented by the reduced negativities disappear (or reappear) for different values of $\kappa /\alpha$ as we compare them with their counterparts from Fig.\ref{fig8}b, and other types of the entangled states are generated (see Tab.\ref{tab2}b). We see that our system is able to produce various types of tripartite entanglement, and is fragile on the values of the parameters describing it. So, by the appropriate tuning of the parameters we can switch the final state of the system from one kind of the state to another. Moreover, what is interesting, for the weak damping case tripartite negativity can increase with growing value of $\kappa /\alpha$. When we observe this feature bipartite entanglement is absent -- reduced negativities are equal to zero. Moreover, what is seen from Figs \ref{fig8}b and \ref{fig8}c, there exists for the both cases some range of the values of damping constant for which final state of the system is not entangled -- all the negativities describing tri- and bipartite entanglement are equal to zero.

\begin{table}[h]
\begin{center}
\raisebox{3cm}{\textbf{a)}}
\begin{tabular}{|c|c|}
\hline
damping parameter & type of tripartite entanglement\\
\hline
$\kappa < 0.4\alpha$ & III-1 \\
\hline
$0.4\alpha<\kappa<4.5\alpha$ & III-3 \\
\hline
$4.5\alpha<\kappa<4.8\alpha$ & III-2 \\
\hline
$4.8\alpha<\kappa<5.1\alpha$ & --- \\
\hline
$5.1\alpha<\kappa<5.4\alpha$ & III-2 \\
\hline
$\kappa >5.4\alpha$ & III-3 \\
\hline
\end{tabular}\\
\raisebox{3cm}{\textbf{b)}}
\begin{tabular}{|c|c|}
\hline
damping parameter & type of tripartite entanglement\\
\hline
$\kappa < 0.65\alpha$ & III-0 \\
\hline
$0.65\alpha<\kappa<0.82\alpha$ & III-2 \\
\hline
$0.82\alpha<\kappa<3.14\alpha$ & III-3 \\
\hline
$3.14\alpha<\kappa<3.53\alpha$ & III-2\\
\hline
$3.53\alpha<\kappa<3.94\alpha$ & - \\
\hline
$3.94\alpha<\kappa<4.67\alpha$ & III-2 \\
\hline
$\kappa>4.67$ & III-3 \\
\hline
\end{tabular}
\caption{Different types of full tripartite entanglement generated for long-time limit for the cases when ${\varepsilon}=\frac{12 \alpha}{10 - \sqrt{28}}$ (a), and ${\varepsilon}=\frac{12 \alpha}{10 - \sqrt{28}}+\Delta$ where $\Delta =-0.6\alpha$ (b).\label{tab2}}
\end{center}
\end{table}

\subsection{Phase damping}
\label{phase_damp}

In this section we consider the influence of phase damping on the generation of various tripartite entangled states. When we include phase damping effects, the system does not loose its energy and populations of the states represented by
diagonal matrix elements do not decay. For such cases, decoherence effects and, hence, entanglement losses can be observed. This is an effect of decay of other than diagonal matrix elements. Decoherence induced by the phase-damping reservoir is related to random changes in the relative phases of superposed states during the system's time-evolution. Moreover, dephasing processes can lead  to other interesting phenomena such as sudden death of entanglement  or/and its reappearing.
 
To describe phase damping effects we apply the following master equation~\cite{WM08, GZ00}:
\begin{eqnarray}
\label{mastereq2}
\frac{d\hat{\rho}}{dt}=&&-\frac{1}{i}\left(\hat{\rho}\hat{H}-\hat{H}\hat{\rho}\right) \\
&&+\sum\limits_{j=1}^{3} \frac{\kappa_{j}}{2}\left[ 2\hat{a}^\dagger_{j} \hat{a}_{j} \hat{\rho} \hat{a}^\dagger_{j} \hat{a}_{j} - \left( \hat{a}^\dagger_{j} \hat{a}_{j} \right)^2 \hat{\rho} - \hat{\rho}  \left( \hat{a}^\dagger_{j} \hat{a}_{j} \right)^2 \right], \nonumber
\end{eqnarray}
where $\kappa_{j}$ ($j=\{1,2,3\}$) is a damping parameter corresponding to the modes $1$, $2$ and $3$, respectively. Analogously, as for the cases discussed in the previous section where the system was amplitude damped,  we assume  here that damping parameters are the same for all modes $\kappa_1=\kappa_2=\kappa_3=\kappa$ and are equal to $0.1\alpha$, and the excitation strength $\alpha=0.001\chi$. Moreover, we consider two cases when the internal coupling parameter ${\varepsilon}=\frac{12 \alpha}{10 - \sqrt{28}}$ and ${\varepsilon}=\frac{12 \alpha}{10 - \sqrt{28}}+\Delta$ ($\Delta =-0.6\alpha$), as previously.

\begin{figure}[h]
\begin{center}
\raisebox{6cm}{\textbf{a)}}
\includegraphics[width=0.5\textwidth]{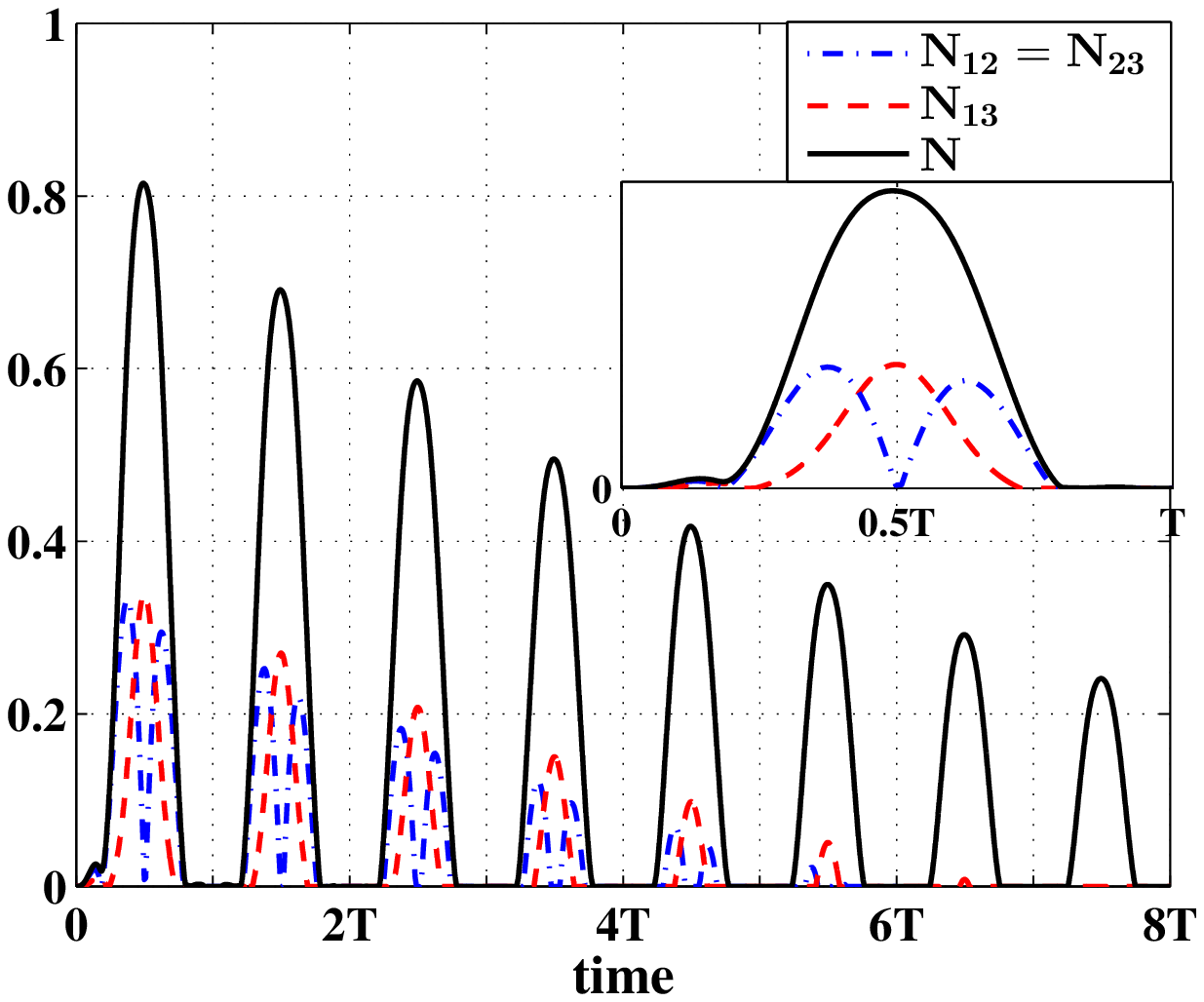}\\
\raisebox{6cm}{\textbf{b)}}
\includegraphics[width=0.5\textwidth]{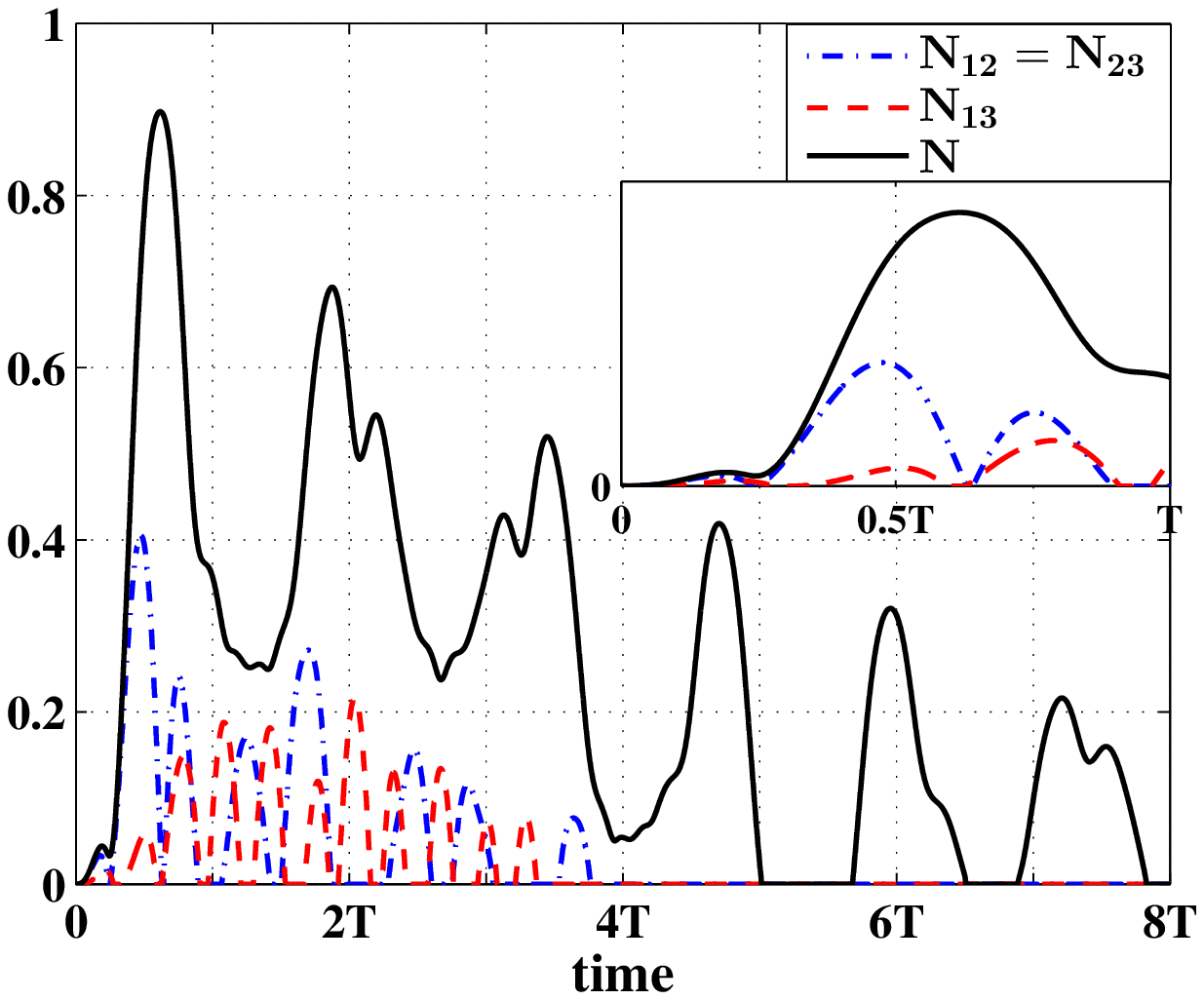}
\end{center}
\caption{Time evolution of the reduced and tripartite negativities for ${\varepsilon}=\frac{12 \alpha}{10 - \sqrt{28}}$ (a) and ${\varepsilon}=\frac{12 \alpha}{10 - \sqrt{28}}+\Delta$ where $\Delta =-0.6\alpha$ (b). Time is measured in units of  $T=\frac{\sqrt{5-\sqrt{7}}\pi}{2\alpha}$,  phase damping parameter $\kappa=0.1\alpha$ and $\alpha=0.001\chi$.}
\label{fig9}
\end{figure}

Thus, Fig.\ref{fig9}a shows time-evolution of the all negativities considered here when ${\varepsilon}=\frac{12 \alpha}{10 - \sqrt{28}}+\Delta$ ($\Delta =-0.6\alpha$) (the system evolves periodically). It is seen, that thanks to the presence of the phase damping, the both effects: sudden death of entanglement and entanglement revival are present in the system. These two phenomena can be observed for the bipartie and tripartite entanglement, as well. 
In a long-time limit all negativities disappear, so the entanglement of any kind does not survive, contrary to the case of the amplitude damping discussed in the previous section (see non-zero values of the negativities for the steady state in Fig.\ref{fig8}). In this limit all eight states involved in the system's evolution are equally populated with the same probability equal to $0.125$.

Nevertheless, for shorter times considerable amount of the tripartite entanglement can be generated. For $t=T/2$, tripartite negativity $N$ reaches its greatest maximum and slightly exceeds $0.8$. At the same moment of time, reduced negativity $N_{13}$ becomes maximal, whereas the negativities $N_{12}=N_{23}$ describing entanglement between two neighbouring oscillators (modes) becomes equal to zero. For such  situation we generate the III-1 class state $\vert 0 \rangle_{2}\pm\vert 1 \rangle_{2})(\vert 00\rangle_{13}\pm\vert 11\rangle_{13})$ which is slightly perturbed by $\vert 001\rangle$, $\vert 100\rangle$, $\vert 011\rangle$, $\vert 110\rangle$. Thanks to the periodic evolution of the system, at the next maxima of $N$ we see the similar situation, although the degree of the entanglement becomes smaller and smaller as we reach successive maxima. Additionally, when all the reduced negativities describing bipartite entanglement differ from zero and $N\neq 0$ as well, W type states are generated. 
When we decrease internal coupling (${\varepsilon}=\frac{12 \alpha}{10 - \sqrt{28}}+\Delta$ where $\Delta =-0.6\alpha$) we observe a similar behaviour of our system, albeit some irregularities appear in the time-evolution of the negativities. Moreover, analogously to the situation depicted in Fig.\ref{fig7} maximal value of $N$ describing tripartite entanglement increases and becomes equal to $\simeq 0.9$. What is interesting, all reduced negativities $N_{ij}$ ($\{i,j\}=\{1,2,3\}$) after some period of time when they exhibit irregular oscillations (for $t>\sim 4T$), becomes equal to zero, and we observe sudden death of bipartite entanglement without its revival.

\section{Summary}
\label{summary}

We have discussed here the model of a chain of three nonlinear oscillators excited by two external coherent fields. Especially, we were interested in a time-evolution of the quantum correlations present in the system, and the possibility of generation of the entangled quantum states which are especially interesting from the point of view of quantum information theory.  As the model consists of the three oscillators, we have discussed the both: bi- and tripartite entanglement.

We have shown that under some conditions our model can be treated as \textit{nonlinear quantum scissors} and evolves within a set of eight states $|ijk\rangle$, where $\{i,j,k\}=\{0,1\}$. In consequence, the system considered here can be treated as  a 3-qubit system. 
We have shown that system can be a source of entangled 2-qubit and 3-qubit states, including maximally entangled ones. For the case of the 2-qubit entanglement, one can observe not only its flow  between the pairs of oscillators, but also the entanglement can be created even between two oscillators which are not directly coupled together. We have also discussed the connections between the first and second order correlation functions, and various relations between those correlations and creation of the entanglement.

The possibility of creation of tripartite entanglement in our model was discussed, as well. It has been shown that it is possible to obtain almost maximally entangled 3-qubit state if only the pair of externally excited oscillators is entangled, and we observe only second order correlations between the modes corresponding to these two oscillators. Moreover, the states of W-type are created if all the 2-qubit subsystems are entangled and $g^{(2)}$ functions indicate either correlations for all subsystems or anticorrelations for all of them. We have shown that it is also possible to generate the state which is very close to the GHZ state.

We have also considered long-time solutions and discussed their dependence on the strength of damping effects. For the amplitude damping case we can observe sudden death of the bipartite entanglement and its sudden revivals, whereas for the phase damping case, the phenomenon of sudden death can be observed for the both: bi- and tripartite entanglement. 

What is the most important, different classes of tripartite entangled states can be obtained in a stationary state limit, and the type of the final entangled state strongly depends on the values of the parameters describing the system. In consequence, we can ''switch'' the final state of the system by adiabatic tuning these  parameters. Moreover, there is a range of damping constant values for which none of the 3-qubit entangled states (and none of 2-qubit ones) is created in steady state limit. 

We believe that all those facts and the generality of our model allows for the conclusion that physical systems governed by the effective Hamiltonian describing our model can be not only a potential source of various bi- and tripartite entangled states, but also interesting finding concerning various types of quantum correlations and relations among them.

\section{Acknowledgement}
\label{Acknowledgement}
 W.L. acknowledges Vietnam Ministry of Education
Grant No. B2014-42-29 for support. W.L. would like to thank Prof. Ryszard Horodecki for his stimulating questions and discussion concerning sudden death of entanglement for tripartite systems.  


%

\end{document}